\documentclass[iop,apjl]{emulateapj}
\usepackage{natbib}
\usepackage{graphicx}
\usepackage{epsfig}
\usepackage{amsmath}
\usepackage{rotating}

\newcommand{\aotsy}{A370}
\newcommand{\asen}{A1689}
\newcommand{\aetf}{Abell 1835}

\newcommand{\attnt}{A2219}
\newcommand{\attny}{Abell 2390}

\newcommand{\rxjthirteen}{RX J$\,1347{-}1145$}

\def \myfigurewidth {0.45}

\shorttitle{Zemcov et al.}
\shortauthors{Zemcov et al.}

\begin{document}

\slugcomment{Submitted to ApJL 2013 April 3; accpeted 2013 April 26;
  published 2013 May 15.}

\title{HerMES: A DEFICIT IN THE SURFACE BRIGHTNESS OF THE COSMIC
  INFRARED BACKGROUND DUE TO GALAXY CLUSTER GRAVITATIONAL LENSING$^{\ast}$}

\author{M.~Zemcov\altaffilmark{1,2},
A.~Blain\altaffilmark{3},
A.~Cooray\altaffilmark{1,4},
M.~B{\'e}thermin\altaffilmark{5,6},
J.~Bock\altaffilmark{1,2},
D.L.~Clements\altaffilmark{7},
A.~Conley\altaffilmark{8},
L.~Conversi\altaffilmark{9},
C.D.~Dowell\altaffilmark{1,2},
D.~Farrah\altaffilmark{10,11},
J.~Glenn\altaffilmark{8,12},
M.~Griffin\altaffilmark{13},
M.~Halpern\altaffilmark{14},
E.~Jullo\altaffilmark{15},
\mbox{J.-P.}~Kneib\altaffilmark{15,16},
G.~Marsden\altaffilmark{14},
H.T.~Nguyen\altaffilmark{1,2},
S.J.~Oliver\altaffilmark{10},
J.~Richard\altaffilmark{17,18},
I.G.~Roseboom\altaffilmark{10,19},
B.~Schulz\altaffilmark{1,20},
Douglas~Scott\altaffilmark{14},
D.L.~Shupe\altaffilmark{1,20},
A.J.~Smith\altaffilmark{10},
I.~Valtchanov\altaffilmark{9},
M.~Viero\altaffilmark{1},
L.~Wang\altaffilmark{10}, and
J.~Wardlow\altaffilmark{4}}
\altaffiltext{$\ast$}{Herschel is an ESA space observatory with science instruments provided by European-led Principal Investigator consortia and with important participation from NASA.}
\altaffiltext{1}{California Institute of Technology, 1200 East
  California Boulevard., Pasadena, CA 91125, USA; zemcov@caltech.edu}
\altaffiltext{2}{Jet Propulsion Laboratory, 4800 Oak Grove Drive,
  Pasadena, CA 91109, USA}
\altaffiltext{3}{Physics \& Astronomy, University of Leicester, University Road, Leicester, LE1 7RH, UK}
\altaffiltext{4}{Department of Physics \& Astronomy, University of
  California, Irvine, CA 92697, USA}
\altaffiltext{5}{Laboratoire AIM-Paris-Saclay, CEA/DSM/Irfu-CNRS-Universit\'e Paris Diderot, CE-Saclay, pt courrier 131, F-91191 Gif-sur-Yvette, France}
\altaffiltext{6}{Institut d'Astrophysique Spatiale (IAS), b\^atiment 121, Universit\'e Paris-Sud 11 and CNRS (UMR 8617), F-91405 Orsay, France}
\altaffiltext{7}{Astrophysics Group, Imperial College London, Blackett Laboratory, Prince Consort Road, London SW7 2AZ, UK}
\altaffiltext{8}{Center for Astrophysics and Space Astronomy 389-UCB,
  University of Colorado, Boulder, CO 80309, USA}
\altaffiltext{9}{Herschel Science Centre, European Space Astronomy Centre, Villanueva de la Ca\~nada, E-28691 Madrid, Spain}
\altaffiltext{10}{Astronomy Centre, Deptartment of Physics \& Astronomy, University of Sussex, Brighton BN1 9QH, UK}
\altaffiltext{11}{Department of Physics, Virginia Tech, Blacksburg, VA
  24061, USA}
\altaffiltext{12}{Deptartment of Astrophysical and Planetary Sciences,
  CASA 389-UCB, University of Colorado, Boulder, CO 80309, USA}
\altaffiltext{13}{School of Physics and Astronomy, Cardiff University, Queens Buildings, The Parade, Cardiff CF24 3AA, UK}
\altaffiltext{14}{Department of Physics \& Astronomy, University of British Columbia, 6224 Agricultural Road, Vancouver, BC V6T~1Z1, Canada}
\altaffiltext{15}{Aix-Marseille Universit\'e, CNRS, LAM (Laboratoire
  d'Astrophysique de Marseille) UMR7326, F-13388 Marseille, France}
\altaffiltext{16}{Laboratoire d'Astrophysique, Ecole Polytechnique Federale de Lausanne (EPFL), Observatoire de Sauverny, CH-1290 Versoix, Switzerland}

\setcounter{footnote}{0}

\begin{abstract}
  We have observed four massive galaxy clusters with the SPIRE
  instrument on the \textit{Herschel Space Observatory} and measure a
  deficit of surface brightness within their central region after
  removing detected sources.  We simulate the effects of instrumental
  sensitivity and resolution, the source population, and the lensing
  effect of the clusters to estimate the shape and amplitude of the
  deficit.  The amplitude of the central deficit is a strong function
  of the surface density and flux distribution of the background
  sources.  We find that for the current best fitting faint end number
  counts, and excellent lensing models, the most likely amplitude of
  the central deficit is the full intensity of the cosmic infrared
  background (CIB).  Our measurement leads to a lower limit to the
  integrated total intensity of the CIB of $I_{250 \, \mu \mathrm{m}}
  >0.69_{-0.03}^{+0.03} (\mathrm{stat.})  _{-0.06}^{+0.11}
  (\mathrm{sys.})\,$MJy sr$^{-1}$, with more CIB possible from both
  low-redshift sources and from sources within the target clusters.
  It should be possible to observe this effect in existing high
  angular resolution data at other wavelengths where the CIB is
  bright, which would allow tests of models of the faint source
  component of the CIB.
\end{abstract}

\maketitle


\vspace*{0.2cm}
\section{Introduction}
\label{S:intro}

\footnotetext[17]{Centre de Recherche Astronomique de Lyon,
  Universit\'e Lyon 1, 9 avenue Charles Andr\'e, F-69230 Saint-Genis
 Laval, France}
\footnotetext[18]{CNRS, UMR 5574, Ecole Normale Sup\'erieure de Lyon,
  F-69007 Lyon, France}
\footnotetext[19]{Institute for Astronomy, University of Edinburgh, Royal Observatory, Blackford Hill, Edinburgh EH9 3HJ, UK}
\footnotetext[20]{Infrared Processing and Analysis Center, MS 100-22,
  California Institute of Technology, JPL, Pasadena, CA 91125, USA}
\setcounter{footnote}{0}

The effect of gravitational lensing is to redistribute the intensity
from sources behind the lens into images with different positions and
brightnesses, while conserving surface brightness
\citep{Schneider1992}.  This means that gravitational lensing not only
magnifies the background sources, but also changes their apparent
density on the sky.  The details of whether the number counts of
background sources seen through a foreground gravitational lens are
increased or decreased depend on the properties of the lens and the
slope of the faint counts \citep{Refregier1997}, a process known as
magnification bias \citep{Turner1980}.

In this Letter, we report detection of a deficit in the surface
brightness of the cosmic infrared background (CIB) in the centers of
massive galaxy clusters measured using the SPIRE instrument
\citep{Griffin2010} on the \textit{Herschel Space Observatory}
\citep{Pilbratt2010}.  To interpret these observations and understand
the consequences of lensing a background field, the lensing properties
of the cluster and the background source population are carefully
simulated in a large number of realizations.  We concentrate on the
intensity profile after removing detected sources to highlight faint
fluctuations in the CIB. The resulting effect after such a removal is
a localized region of decreased surface brightness at the cluster
center.  We use these observations to constrain the intensity of the
submillimeter (sub-mm) background at $250 \, \mu$m.

\section{Observation of the Deficit}
\label{S:observation}

This study uses confusion-limited maps of galaxy clusters from
the \textit{Herschel} Multi-tiered Extragalactic Survey (HerMES;
\citealt{Oliver2012}).  The SPIRE data are reduced using a
combination of the {\sc hipe} \citep{Ott2006} and {\sc smap}
(\citealt{Levenson2010}; \citealt{Viero2012}) packages.  The
sample of clusters is listed in \citet{Oliver2012} and M.~Zemcov et
al.~(in preparation).

In order to study the strongly lensed regions it is necessary to
restrict our attention to those clusters large enough that
$\Omega_{\mathrm{c}} > 3 \Omega_{\mathrm{s}}$, where
$\Omega_{\mathrm{c}}$ is the solid angle of the negative magnification
region and $\Omega_{\mathrm{s}}$ is the solid angle of the SPIRE
beam. This cut leaves only four clusters from the HerMES sample of
twelve: \aotsy, \asen, \attnt\ and \rxjthirteen.  The negative
magnification region in the image plane maps to the interior of the
lens caustic in the source plane, so it is a reasonable tracer of the
area which undergoes strong lensing.  The mapping from total mass to
lensing characteristics is complex and the area $\Omega_{\mathrm{c}}$
is not strictly proportional to cluster mass; however, this cut does
effectively restrict our analysis to the most massive
centrally condensed systems.

\begin{figure}[ht]
\centering
\epsfig{file=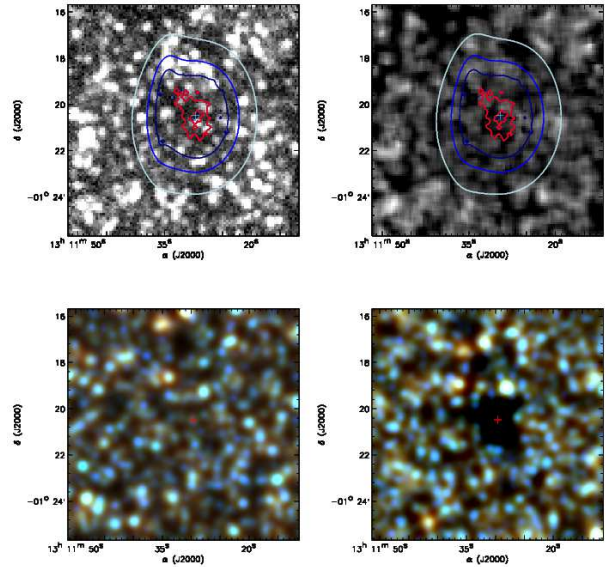,width=\myfigurewidth\textwidth}
\caption{Lensed and field images of the CIB from SPIRE.  The upper
  left-hand panel shows the $250 \, \mu$m \asen\ image, and the upper
  right-hand panel shows the same image after detected sources have
  been subtracted.  Because the detection threshold is equal to the
  confusion limit and the noise is not uniform over the image, at the
  high contrast level of the right hand image $\sim 1
  \sigma_{\mathrm{conf}}$ noise fluctuations are visible towards the
  edges of the map.  The colored contours show the iso-magnification
  levels from the gravitational lensing model for a source at $z =1.5$
  with $\mu < 0$ enclosed by the central red contours and $\mu =
  \{3,2,1.5\}$ in lightening shades of blue.  The lower two panels
  show SPIRE three-band false color simulations of the CIB to $S \lesssim
  100 \,$nJy, not including noise.  The lower left-hand panel is a
  image of the CIB without lensing, while the right-hand panel shows
  the same background that has been propagated through a lens model
  for \asen; no emission from the cluster or SZ effect is included.}
\label{fig:maps}
\end{figure}

\begin{figure*}[ht]
\centering
\epsfig{file=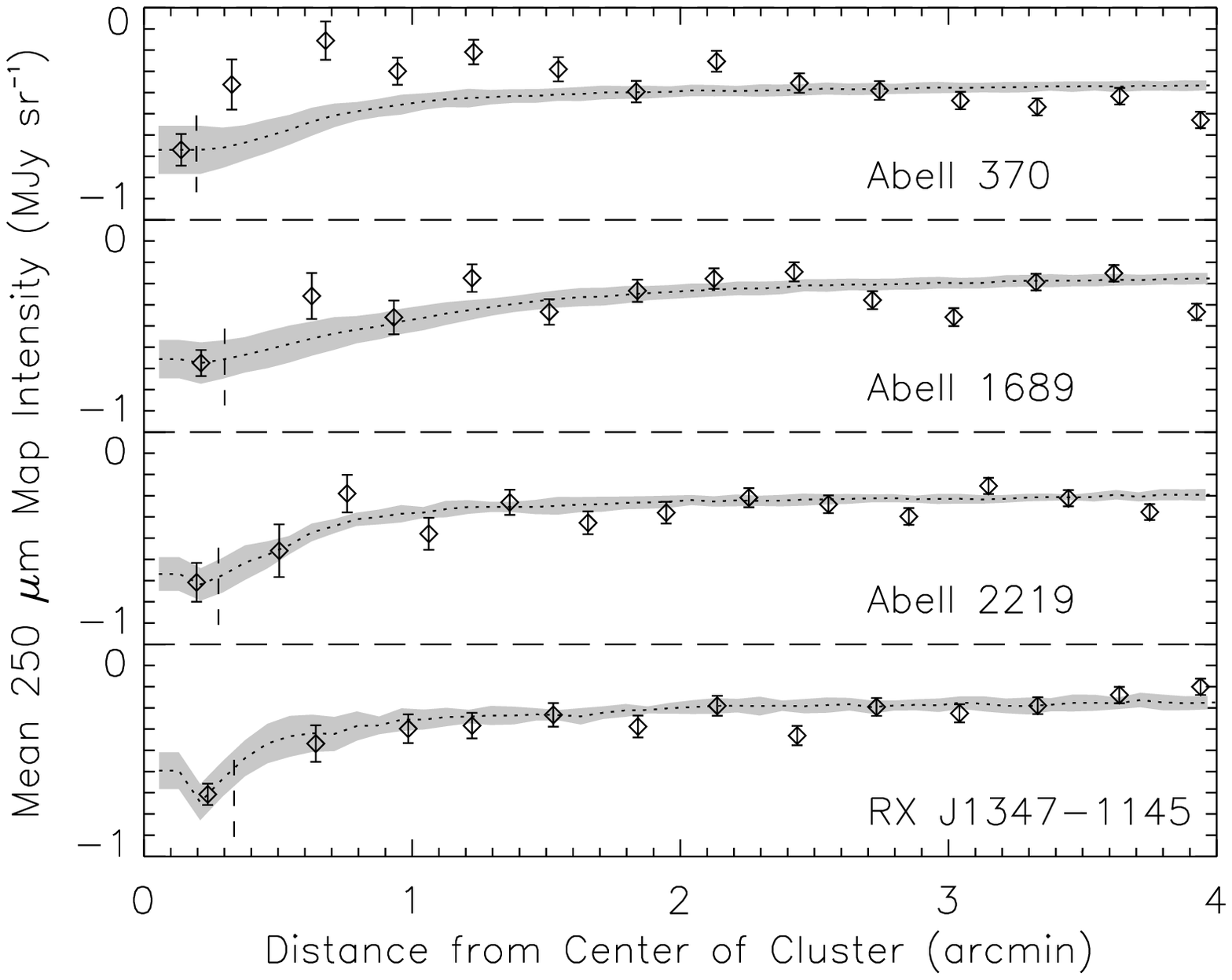,width=0.45\textwidth}
\epsfig{file=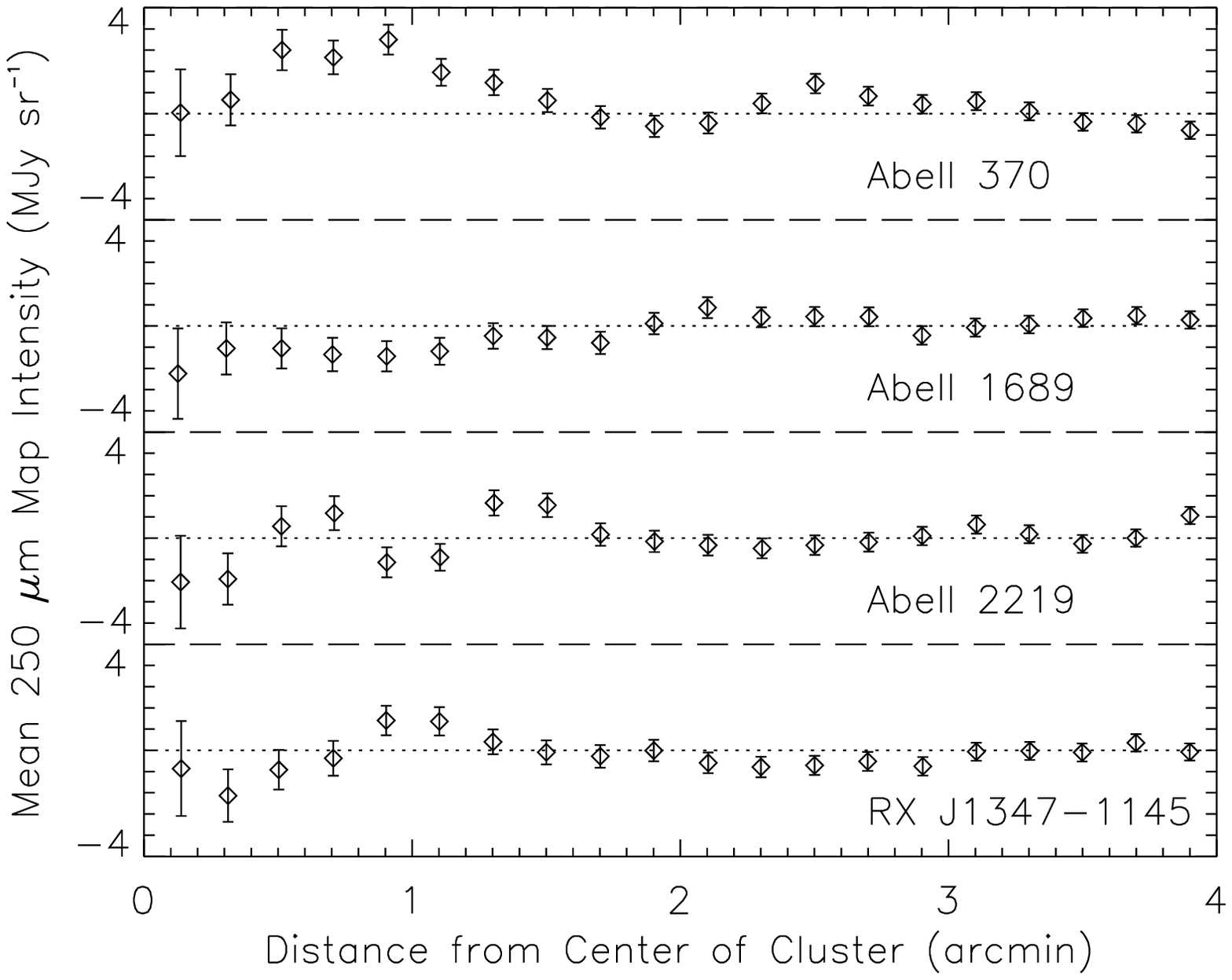,width=0.45\textwidth}
\caption{Intensity profiles towards four clusters, illustrating
  gravitational lensing of the CIB. The left hand plots show the mean
  flux density in $0^{\prime}.25$-wide annular bins for each cluster
  after all detected sources have been removed. The central data point
  is constructed from averaging map pixels within the effective radius
  characteristic of the clusters' critical lines (marked by the
  vertical dashed lines). At larger radii, the data points represent
  uncertainty-weighted averages starting at the characteristic
  critical line radius.  The mean level of each simulated map is
  constructed to be zero before source subtraction and is reduced by
  $\sim 0.3 \,$MJy sr$^{-1}$ after a large fraction of the CIB has
  been detected as sources and removed; the offset varies between
  targets depending on the details of the catalog.  The uncertainties
  on the data points reflect the photometric accuracy of the
  measurements rather than the precision with which the mean should
  differ from the model.  Dotted lines and grey contours show the mean
  and standard deviation of our sky model, calculated from simulations
  of the CIB lensed by each cluster and passed though our data
  analysis pipeline.  The right hand plots show the annular averages
  of the raw cluster images with no source extraction, which highlight
  the structure of the surface brightness which is subtracted by
  removing detected sources. }
\label{fig:clustercenter}
\end{figure*}

To measure the deficit, a catalog is generated for each cluster at
each SPIRE wavelength using the {\sc scat} algorithm \citep{Smith2012}
based on $250 \, \mu$m source-selection.  Our goal is not to identify
particular point sources, but rather to remove emission from bright
point sources, so employing a low detection threshold is reasonable.
We chose a $1 \sigma$ threshold, where $\sigma$ is the map
root-mean-square variation, dominated in these maps by confusion
noise.  The measurement is not sensitive to the precise value of this
number provided it is not so high as to leave residual sources ($> 3
\sigma$) nor so low as to needlessly mask a large fraction of the map
($< 0.5 \sigma$).  This yields a catalog of $\sim 500$ sources in each
cluster field which are then subtracted from the cluster image using a
Gaussian model of the beam.  An example of the resulting image for one
cluster is shown in Figure \ref{fig:maps}, which shows a statistical
deficit in surface brightness near the center of the cluster.  A
simple way to visualize this deficit is to plot annular averages of
the source-subtracted maps, as shown in Figure \ref{fig:clustercenter}.
Similar averages centered on randomly chosen positions in SPIRE images
do not show this deficit.

\section{Theoretical Grounds}
\label{S:theory}

Gravitational lensing conserves surface brightness and therefore does
not alter the mean intensity of the extragalactic background light at
any wavelength.  However, this statement applies only when all sources
are kept in calculating the intensity profile and when averaged over
many directions through the cluster -- for some configurations of the
background sky unusually bright images can be found. The measurement
reported here involves the removal of bright detected sources, which
leads to a situation where lensing shows a profile that does not
conserve intensity.  This does not imply that the total number of
photons and therefore intensity is not conserved, as discussed in
\citet{Refregier1997}, since detected sources are removed.

When viewed through a lensing cluster imposing a magnification $\mu$
on background sources, a background source with flux density $S$ is
magnified such that the observed flux density is $S^{\prime} = \mu
S$. The resulting increase in the flux density is accompanied by a
corresponding decrease in the projected surface density of galaxies;
the observed number density of sources through the lens is modified to
be $N^{\prime} = N/\mu$.  The overall effect of this is to modulate
existing fluctuations in the background source field.  Using the
parameterization that the intrinsic number count of the faint
background sources scales as $dN/dS \propto S^{-\Gamma}$, the
differential number counts lensed through the cluster become
$dN^{\prime}/dS^{\prime} \propto \mu^{\Gamma-2} S^{-\Gamma}$.  The
differential counts imaged through the cluster are either decreased or
increased depending on whether the intrinsic counts have a slope
$\Gamma$ smaller than or greater than 2.  If sources are not removed
from the image, the total CIB surface brightness is conserved
\citet{Refregier1997}.

The observations reported here do not constrain the exact number
counts of sources or the difference of the counts of sources through
and away from the cluster.  One reason for this is that the
observations are limited by the angular resolution of the
SPIRE instrument, which results in a blending of the faintest
sources. The effects of limited-resolution observations of the
intensity profile of background sources through the cluster are
described in \citet{Blain1997} and \citet{Blain2002a}.  High resolution
observations with ALMA will allow more precise measurements of the
difference in the counts away from and through the cluster.

\section{Numerical Simulation}
\label{sS:sims}

We can use detailed lensing models of these well-studied clusters and
models of the source counts and redshift distributions of the sources
which comprise the CIB to generate numerical simulations of the
observed sky brightness, and explore how the amplitude of an observed
deficit depends on source model parameters.  To simulate the sub-mm
background, we use the model described in \citet{Bethermin2011}, which
is tuned to match a variety of observed number counts $dN/dS$ and is
forced to integrate to the total intensity of the CIB
\citep{Lagache2000}.  The model associates both a spectral shape and
redshift to each source, down to $z=0$.  These simulations do not
include coherent clustering of the galaxies which comprise the sub-mm
background, which is a small effect on the scales of interest here
($\theta < 3$ arcmin; \citealt{Viero2012}).  Poisson noise is
modeled by the simulation and accounts for the cosmic variance one
would expect in the real sky. The simulated images have area $0.25$
deg$^{2}$ and contain $\sim 5 \times 10^{4}$ sources to source fluxes
$S_{\mathrm{min}} = 50 \,$nJy. 

These simulated backgrounds are then lensed using {\sc lenstool}
models (\citealt{Kneib1996}; \citealt{Jullo2007};
\citealt{Jullo2009}), which are built using optical measurements of
strong lensing in the cluster fields (\aotsy\ --
\citealt{Richard2010a}; \asen\ -- \citealt{Limousin2007}; \attnt\ --
\citealt{Smith2005} \citealt{Richard2010b}; \rxjthirteen\ --
\citealt{Bradac2008}), modeling all of the known high-magnification
images measured for each cluster.  The {\sc lenstool} models replicate
all of the phenomenology arising from the complex dark-matter
potentials of each cluster, including multiple images, giant arcs,
etc.  Gravitational lensing is achromatic, so the same lens model
applies at all wavelengths.  Sources with $z < z_{_\mathrm{c}}$ are
included in the simulated sky map, but are not lensed; all the sources
with $z > z_{\mathrm{c}}$ are lensed from redshift planes discretized
into $\delta z =0.1$ steps.  Varying the redshift distribution of the
\citet{Bethermin2011} model within reasonable limits does not change
the results of these simulations.  An example of one such simulation
is shown in the bottom panels of Figure \ref{fig:maps}.  The central
surface brightness deficit is evident in the modeled image in all
three SPIRE bands.

These simulations show that three factors contribute to the effect
reported here: (1) the sizes and typical lensing amplification factors
of clusters, (2) the number densities and redshifts of the sources
responsible for the CIB, and (3) the sensitivity and beam size of
SPIRE.  These combine so that when we examine the lensed cores of
SPIRE images of cluster fields and remove all reasonably significant
sources, we are left with deficits in the surface brightness of the
sky in small regions near the cluster center.

In order to compare the data to the simulations, we use the same
procedure discussed above on simulations of the cluster targets.  In
these simulations, instrumental noise with the same amplitude as the
data is inserted into the simulated, lensed cluster images which are
propagated through an identical detection/subtraction procedure as the
real data, resulting in the gray bands in Figure \ref{fig:clustercenter}
which represent the mean and standard deviation of the model results
drawn from 100 realizations of the background sky.  These capture the
variance of the lensing deficit due to different background
configurations and reflect the range of possible deficit shapes and
amplitudes arising from our ignorance of the exact configuration of
the background sources.

A potential concern is whether simulating sources to $S \sim 50 \,$nJy
is sufficient for the modeling result to converge.  To check this, we
have performed simulations where sources are excluded below
$S_{\mathrm{min}}$.  By $S_{\mathrm{min}} = 10 \, \mu$Jy the map is
densely populated and the effect has converged, and including fainter
sources does not have an appreciable effect on the amplitude of the
deficit.  This is because sources below $10 \, \mu$Jy are rarely
boosted above the $1 \sigma$ detection threshold so do not create a
net deficit after source removal.

\begin{figure}[ht]
\centering
\epsfig{file=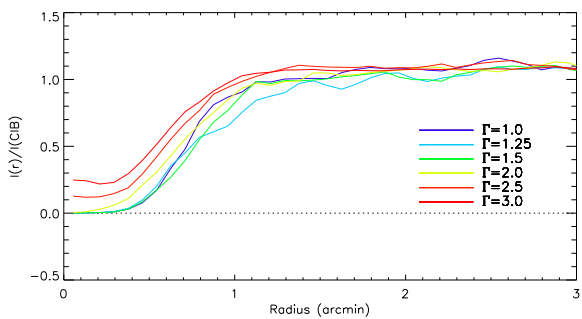,width=\myfigurewidth\textwidth}
\caption{Effect of the slope of the faint end source counts $\Gamma$
  on the lensing deficit.  The surface brightness of the sky is
  averaged into annular bins centered on the cluster and expressed as
  a ratio of the brightness at radius $r$ to the overall sky surface
  brightness.  For slopes similar to that inferred from other
  measurements ($\Gamma \approx 1.5$), the center of our modeled
  clusters has an expectation value of zero; as the slope is
  increased, the probability of a source falling into the caustic
  region is increased until the deficit begins to be filled in.}
\label{fig:simgamma}
\end{figure}

To investigate the dependence of the deficit on the background number
counts, we perform simulations drawn from the measured counts
presented in \citet{Glenn2010}, similar to the calculations in
\citet{Ford2012}.  The slope of the faint end for $0.1 \,$mJy $< S < 2 \,$mJy
is varied in $1.0 \leq \Gamma \leq 3.0$, bracketing the nominal value
$\Gamma = 1.65$ which accounts for the FIRAS background
(\citealt{Fixsen1998}; \citealt{Lagache2000}).  As can be seen in
Figure \ref{fig:simgamma}, for shallow slopes, the surface density of
sources is small enough to produce zero flux in the center of
clusters.  For steeper slopes, the probability that a source falls
behind the center of the cluster image is no longer small and the
surface brightness increases.  Existing constraints on the brightness
of the CIB at $250 \, \mu$m exclude $\Gamma > 2.3$ at $2 \sigma$ in
this range \citep{Fixsen1998}.

\section{Measurement of the CIB at 250 Microns}
\label{S:CIB}

According to our simulations using the \citet{Bethermin2011} model,
statistically the central region of these clusters' images have zero
surface brightness from lensed sources.  At $250 \, \mu$m, emission
from sources within the central arcminute of clusters tends to be
faint.  Therefore a situation arises where the cluster itself is
“invisible”, the probability of a foreground source coinciding with
the cluster center is small, and the background is lensed by the
cluster, so no sources of emission exist in the center of the
cluster.  If this region has zero surface brightness, as the model
suggests, we can use it as an absolute zero point.

\begin{table*}[ht]
\caption{Lensing Deficit CIB Limits and Error Budget \label{tab:cib}}
\begin{tabular}{l|ccccccc}
\hline
Cluster & $\sum \Omega$ & CIB $I_{250 \, \mu \mathrm{m}}$ & 
Stat. $\delta I_{250 \, \mu \mathrm{m}}$ &
Model  $\delta I_{250 \, \mu \mathrm{m}}$ &
BCG $\delta I_{250 \, \mu \mathrm{m}}$ &
Abs.~Cal.~$\delta I_{250 \, \mu \mathrm{m}}$ &
Total $\delta I_{250 \, \mu \mathrm{m}}$ \\

 & $(\Omega_{\mathrm{s}}^{\mathrm{a}})$ & (MJy sr$^{-1}$) & (MJy sr$^{-1}$) &
(MJy sr$^{-1}$) & (MJy sr$^{-1}$) &
(MJy sr$^{-1}$) & (MJy sr$^{-1}$)  \\ \hline

\aotsy & 3.0 & 0.67 & $\pm 0.08$ & $\pm 0.11$ & $+0.05$ & $\pm
0.03$ &  $-0.14,+0.15$ \\

\asen & 7.5 & 0.68 & $\pm 0.06$ & $\pm 0.09$ & $+0.05$ & $\pm 0.03$ &
$-0.11,+0.12$ \\

\attnt & 3.3 & 0.71 & $\pm 0.09$ & $\pm 0.08$ & $+0.05$ & $\pm 0.04$ &
$-0.13,+0.14$ \\

\rxjthirteen & 11.3 & 0.71 & $\pm 0.05$ & $\pm 0.09$ & $+0.05$ & $\pm
0.04$ & $-0.11,+0.12$ \\ \hline

Total & 25.1 & 0.69 & $\pm 0.03$ & $\pm 0.05$ & $+0.09$ & $\pm 0.04$ &
$-0.07,+0.12$ \\ \hline 

\multicolumn{8}{l}{$^{\mathrm{a}} \, 366^{\prime \prime} / \Omega_{\mathrm{s}}$.}

\end{tabular}
\end{table*}

Of course, if the above assumption is incorrect, any sources of
emission local to the cluster center -- for example, from the
brightest cluster galaxy (BCG) or the Sunyaev-Zeldovich (SZ) effect --
will bias the measurement to lower inferred surface brightness.
Though only a $< 0.1\%$ fractional contamination in the $250 \, \mu$m
SPIRE band, the SZ effect is significant at longer sub-mm wavelengths
(\citealt{Zemcov2007}; \citealt{Zemcov2010}) so we defer reporting the
CIB inferred from the two longer wavelength SPIRE bands to future
work.  Sub-mm emission associated with the central region of target
clusters is known \citep{Rawle2012}; for example, in the HerMES sample
(but excluded from this work) \aetf\ and \attny\ are both known to
host BCGs with sub-mm fluxes of many mJy.  Because of this potential
bias, the lensing deficit yields a lower limit to the CIB.  To check
for contamination from galaxies within the critical region, we use
\textit{Spitzer}-MIPS $24 \, \mu$m data.  No BCGs are obvious in the
SPIRE images of the four clusters, but we can use the $24 \, \mu$m --
$250 \, \mu$m BCG flux density ratio measured by \citet{Rawle2012} to
estimate an additional one-sided, positive-going uncertainty to the
central zero point in each image.  This corresponds to a $1 \sigma$
value of $0.4 \,$mJy per beam at the position of the BCG.  We do not
detect other $24 \, \mu$m sources associated with the cluster in the
central deficit region.  Diffuse dust emission associated with the
intracluster medium has never been detected, but from predictions we
expect this kind of emission to be at least an order of magnitude
smaller than the surface brightness of the lensing deficit at $250 \,
\mu$m \citep{Popescu2000}.

To determine the absolute intensity of the CIB, the average of each
map in its central region is used to generate a zero point for the
image.  The mean brightness of the resulting image, excluding the
central region, is then computed.  To estimate the uncertainty
associated with each measurement, we compute the quadrature sum of the
statistical uncertainty of the maps as traced by the standard
deviation of the pixels in the central region (Stat.~$\delta I_{250 \,
  \mu \mathrm{m}}$), the one-sided uncertainty from the BCG emission
(BCG~$\delta I_{250 \, \mu \mathrm{m}}$), the uncertainty from the
simulations associated with the configuration of the background source
(Model $\delta I_{250 \, \mu \mathrm{m}}$), and the $5 \,$\% absolute
calibration uncertainty of SPIRE (Abs.~Cal.~$\delta I_{250 \, \mu
  \mathrm{m}}$).  We estimate the Eddington bias associated with the
effect of subtracting sources from the maps before computing the mean
of the central region by comparing simulations including and excluding
simulated noise, resulting in an estimate of $\delta I < 0.03 \,$MJy
sr$^{-1}$ for the Eddington bias on the sample.  Table \ref{tab:cib}
lists the areas of negative magnification in each target, the inferred
CIB brightness for each of the four cluster fields that pass all of
the selection cuts, and the uncertainty budget associated with these
measurements.  The statistical-uncertainty weighted mean of this
value, $0.69_{-0.07}^{+0.12} \,$MJy sr$^{-1}$, is the inferred
brightness of the CIB at $250 \, \mu$m from the lensing deficit
method.  This value is consistent with the FIRAS values determined by
\citet{Puget1996}, \citet{Fixsen1998}, and \citet{Lagache2000} in the
same band within $1 \sigma$.

\section{Discussion}
\label{S:discussion}

We have detected an interesting new phenomenon which is due to the
properties of the sub-mm background and gravitational lensing in
massive clusters.  This deficit in the measured surface brightness
constrains the smallest allowable surface brightness of the CIB and so
is another way to limit its absolute brightness, adding to a list
which includes $P(D)$ analyses, stacking, and source counting, in
addition to direct photometric measurement.  Furthermore, in principle
the presence of this deficit places interesting constraints on the
faint end of the number counts, although larger samples and
comprehensive modeling are required to convert measurements to source
count constraints.

Because gravitational lensing is achromatic, this effect occurs in all
of the SPIRE bands.  Other instruments working at sub-mm and mm
wavelengths like ACT/SPT and CCAT should also be able to measure this
deficit effect, assuming that the background in question is
sufficiently far behind the lensing cluster, and that the
instrumentation is sensitive to sources at levels similar to SPIRE's
with enough angular resolution to resolve the central region of
clusters.  In addition, this effect produces a complicated, spatially
structured CIB surface brightness distribution which is a potential
foreground for high resolution SZ effect measurements at longer
wavelengths.

\section*{Acknowledgements} 

SPIRE has been developed by a consortium of institutes led
by Cardiff Univ. ~(UK) and including: University of Lethbridge (Canada);
NAOC (China); CEA, LAM (France); IFSI, University of Padua (Italy);
IAC (Spain); Stockholm Observatory (Sweden); Imperial College
London, RAL, UCL-MSSL, UKATC, University of Sussex (UK); Caltech,
JPL, NHSC, University of Colorado (USA). This development has been
supported by national funding agencies: CSA (Canada); NAOC 
(China); CEA, CNES, CNRS (France); ASI (Italy); MCINN (Spain);
SNSB (Sweden); STFC, UKSA (UK); and NASA (USA).

HCSS/HSpot/HIPE are joint developments by the Herschel Science Ground
Segment Consortium, consisting of ESA, the NASA Herschel Science
Center, and the HIFI, PACS and SPIRE consortia.

This work is based in part on archival data obtained with the
\textit{Spitzer Space Telescope}, which is operated by JPL/Caltech under a contract with
NASA. 

Support for this work was provided by NASA.


\end{document}